\documentclass[prl,twocolumn,amsmath,amssymb,amsfont,superscriptaddress,showpacs,lengthcheck]{revtex4}

\usepackage{graphicx}
\usepackage{bm}

\begin{document}

\title{Feedback Cooling of the Normal Modes of a Massive Electromechanical System to Submillikelvin Temperature}

\author{A. Vinante}
\email{vinante@science.unitn.it}
\affiliation{Istituto di Fotonica e Nanotecnologie, CNR-Fondazione Bruno Kessler, 38100, Povo, Trento (Italy)}
\affiliation{INFN, Gruppo Collegato di Trento, Sezione di Padova, 38100, Povo, Trento (Italy)} 

\author{M. Bignotto}
\affiliation{Dipartimento di Fisica, Universit\`{a} di Padova, 35131, Padova (Italy)} 
\affiliation{INFN, Sezione di Padova, 35131, Padova (Italy)} 

\author{M. Bonaldi}
\affiliation{Istituto di Fotonica e Nanotecnologie, CNR-Fondazione Bruno Kessler, 38100, Povo, Trento (Italy)}
\affiliation{INFN, Gruppo Collegato di Trento, Sezione di Padova, 38100, Povo, Trento (Italy)} 

\author{M. Cerdonio}
\affiliation{Dipartimento di Fisica, Universit\`{a} di Padova, 35131, Padova (Italy)} 
\affiliation{INFN, Sezione di Padova, 35131, Padova (Italy)} 

\author{L. Conti}
\affiliation{Dipartimento di Fisica, Universit\`{a} di Padova, 35131, Padova (Italy)} 
\affiliation{INFN, Sezione di Padova, 35131, Padova (Italy)} 

\author{P. Falferi}
\affiliation{Istituto di Fotonica e Nanotecnologie, CNR-Fondazione Bruno Kessler, 38100, Povo, Trento (Italy)}
\affiliation{INFN, Gruppo Collegato di Trento, Sezione di Padova, 38100, Povo, Trento (Italy)} 

\author{N. Liguori}
\affiliation{Dipartimento di Fisica, Universit\`{a} di Padova, 35131, Padova (Italy)} 
\affiliation{INFN, Sezione di Padova, 35131, Padova (Italy)} 

\author{S. Longo}
\affiliation{INFN, Laboratori Nazionali di Legnaro, 35020, Legnaro, Padova (Italy)}

\author{R. Mezzena} 
\affiliation{Dipartimento di Fisica, Universit\`{a} di Trento, 38100, Povo, Trento (Italy)}
\affiliation{INFN, Gruppo Collegato di Trento, Sezione di Padova, 38100, Povo, Trento (Italy)}

\author{A. Ortolan}
\affiliation{INFN, Laboratori Nazionali di Legnaro, 35020, Legnaro, Padova (Italy)}

\author{G.A. Prodi} 
\affiliation{Dipartimento di Fisica, Universit\`{a} di Trento, 38100, Povo, Trento (Italy)}
\affiliation{INFN, Gruppo Collegato di Trento, Sezione di Padova, 38100, Povo, Trento (Italy)}

\author{F. Salemi} 
\affiliation{Dipartimento di Fisica, Universit\`{a} di Trento, 38100, Povo, Trento (Italy)}
\affiliation{INFN, Gruppo Collegato di Trento, Sezione di Padova, 38100, Povo, Trento (Italy)}

\author{L. Taffarello}
\affiliation{INFN, Sezione di Padova, 35131, Padova (Italy)} 

\author{G. Vedovato}
\affiliation{INFN, Sezione di Padova, 35131, Padova (Italy)} 

\author{S. Vitale} 
\affiliation{Dipartimento di Fisica, Universit\`{a} di Trento, 38100, Povo, Trento (Italy)}
\affiliation{INFN, Gruppo Collegato di Trento, Sezione di Padova, 38100, Povo, Trento (Italy)}

\author{J.-P. Zendri}
\affiliation{INFN, Sezione di Padova, 35131, Padova (Italy)} 

\date{\today}


\begin{abstract}

We apply a feedback cooling technique to simultaneously cool the three electromechanical normal modes of the ton-scale resonant-bar gravitational wave detector AURIGA. The measuring system is based on a dc Superconducting Quantum Interference Device (SQUID) amplifier, and the feedback cooling is applied electronically to the input circuit of the SQUID. Starting from a bath temperature of $4.2$ K, we achieve a minimum temperature of $0.17$ mK for the coolest normal mode. The same technique, implemented in a dedicated experiment at subkelvin bath temperature and with a quantum limited SQUID, could allow to approach the quantum ground state of a kilogram-scale mechanical resonator.

\end{abstract}

\maketitle

The prospect of observing quantum behaviour and investigating decoherence in macroscopic mechanical resonators \cite{armour, penrose} requires cooling to ultralow temperatures, such that the thermal energy becomes comparable to the quantum energy. In addition to conventional refrigeration, this goal requires the use of feedback cooling techniques \cite{tombesi}, unless the frequency is as large as several hundred MHz. Various authors have recently reported advances in cooling mechanical resonators, using either optomechanical \cite{arcizetprl,kleckner,poggio,corbitt,metzger,gigan,arcizetnature} or electromechanical \cite{naik,brown} techniques, and implementing either active feedback \cite{arcizetprl,kleckner,poggio,corbitt} or dynamical back-action effects \cite{metzger,naik,gigan,arcizetnature,brown}. These experiments involved mostly nanomechanical or micromechanical resonators, and have demonstrated cooling capability down to mK temperatures \cite{poggio}. Cooling of a gram-scale optical spring resonator to a few mK has been demonstrated as well using the techniques developed for interferometric gravitational wave (GW) detectors \cite{corbitt}. In this Letter, we show that very efficient cooling can be achieved even in much larger systems, exploiting the techniques developed for resonant-mass GW detectors, based on electromechanical transducers coupled to Superconducting Quantum Interference Device (SQUID) sensors. In this case, one can take advantage of the larger quality factor achievable in macroscopic systems with respect to micromechanical ones. As experimental demonstration, we simultaneously cool the three electromechanical normal modes of the ton-scale resonant-bar GW detector AURIGA. Our cooling technique is based on an electronic feedback directly applied to the input circuit of the SQUID.

\begin{figure}[!h]
\includegraphics{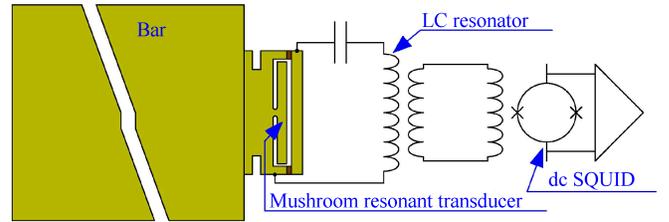}
\caption{(Color online) Scheme of the gravitational wave detector AURIGA. The system comprises three coupled resonators with nearly equal resonant frequency of about 900 Hz: the first longitudinal mode of the cylindrical bar, the first flexural mode of the mushroom shaped resonator, which is also one of the plates of the electrostatic capacitive transducer, and the low-loss electrical LC circuit. The electrical current of the LC resonator is detected by a low noise dc SQUID amplifier.} \label{fig1}
\end{figure}

AURIGA represents the state-of-art in the class of resonant GW detectors \cite{amaldi5}, and has been in continuous operation from year 2004, searching for galactic astrophysical events in collaboration with a world network of detectors \cite{igec2}. It is located in Padua (Italy) and is based on a $2.2 \times 10^3$ kg, $3$ m long bar made of a low loss aluminium alloy (Al5056), cooled to liquid helium temperatures. The fundamental longitudinal mode of the bar, sensitive to gravitational waves, has an effective mass $M$=$1.1 \times 10^3$ kg and a resonance frequency $\omega_B/2\pi=900$ Hz. According to the equipartition theorem, the rms amplitude of the resonator motion is given by $x_{{\rm rms}}  = \left\langle {x^2 } \right\rangle ^{\frac{1}{2}}  = \left( {\frac{{k_B T}}{{M\omega _B ^2 }}} \right)^{\frac{1}{2}}$, where $k_B$ is the Boltzmann constant and $T$ is the temperature. For the AURIGA bar, the rms thermal motion is $x_{\rm rms}=4\times10^{-17}$ m at $T=4.2$ K. The bar resonator motion is detected by a displacement sensor with a sensitivity of order several $10^{-20}$ m/$\sqrt{\rm Hz}$ over a $\sim 100$ Hz bandwidth. This sensitivity is accomplished by a multimode resonant capacitive transducer \cite{3modi} combined with a very low noise dc SQUID amplifier \cite{squid} (Fig.~\ref{fig1}). In this scheme, the bar resonator is coupled to the fundamental flexural mode of a mushroom-shaped lighter resonator, with $6$ kg effective mass and the same resonance frequency. As the mechanical energy is transferred from the bar to the lighter resonator, the motion is magnified by a factor of roughly $15$. A capacitive transducer, biased with a static electric field of $10^7$ V/m, converts the differential motion between bar and mushroom resonator into an electrical current, which is finally detected by a low noise dc SQUID amplifier through a low-loss high-ratio superconducting transformer. The transducer efficiency is further increased by placing the resonance frequency of the electrical LC circuit close to the mechanical resonance frequencies \cite{3modi}, at $930$ Hz. 

\begin{figure}[!h]
\includegraphics{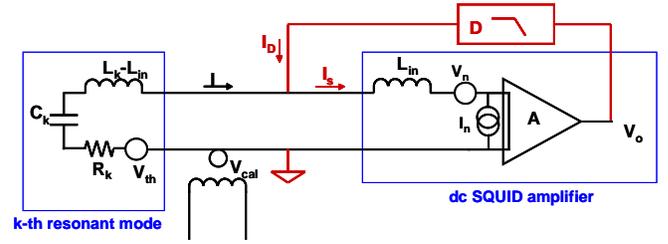}
\caption{(Color online) The feedback cooling scheme. The $k$-th normal mode is approximated, around its resonance frequency, by a series-RLC circuit. In this representation, different modes should be thought of as being in parallel with each other. The dc SQUID is represented as current amplifier. The electronic feedback cooling is obtained by sending back a current $I_D$ phase shifted of $\pi/2$ with respect to $I_s$.} \label{fig2}
\end{figure}

The detector can then be simply modelled as a system of three coupled resonators: its dynamics is described by three normal modes at separate frequencies, each one being a superposition of the bar and transducer mechanical resonators and the LC electrical resonator \cite{amaldi5}. As seen from the SQUID sensor, each mode $k$ ($k$=1,2,3) is modelled as a RLC series electrical mode with an effective inductance $L_k$, capacitance $C_k$ and resistance $R_k$ (Fig.~\ref{fig2}). The total inductance $L_k$ includes also the input inductance $L_{\rm{in}}$ of the SQUID amplifier. Around the resonance frequency of each mode $\omega_k/2\pi$, with $\omega_k=\left(L_k C_k \right)^{-1/2}$, the complex impedance of the circuit is expressed by $Z_k \left( \omega  \right) = R_k  + i\omega L_k  + 1/\left( {i\omega C_k } \right)$. We point out that, although the modes appear as purely electrical as seen from the SQUID, their dynamics actually includes the full motion of both mechanical resonators. In fact, the current in the electrical circuit is linearly related with the mechanical motion of bar and transducer, and the effective impedance parameters $L_k$, $C_k$, $R_k$ of each normal mode are determined by the mechanical and electrical parameters of all resonators \cite{amaldi5}. Thus, cooling the normal modes of the system implies cooling the motion of the mechanical resonators. 
The effective impedance parameters $L_k$, $C_k$, $R_k$  are experimentally estimated by measuring the current $I_s=V_{\rm{cal}}/Z_k$ in response to a calibration voltage signal $V_{\rm{cal}}$, injected through a small calibration coil in series to the SQUID input coil (see Fig.~\ref{fig2}). 
The inductances $L_k$ of  the three modes are respectively $1.66\times 10^{-4}$ H, $1.23\times 10^{-5}$ H and $8.12\times 10^{-6}$ H. The mode resonance frequencies $\omega_k / 2 \pi$ are 865 Hz, 914 Hz, 953 Hz, and the quality factors $Q_k \equiv \omega_k L_k / R_k$  are $1.2 \times 10^6$, $0.88\times 10^6$, and $0.77 \times 10^6$.

The effective temperature $T_k$ of each mode is proportional to the mean square current $\left\langle {I_k ^2 } \right\rangle $ induced by thermal fluctuations driving the mode, according to the equipartition theorem:
\begin{equation}
\left\langle {I_k ^2 } \right\rangle  = \frac{{k_B T_k}}{{L_k }}. \label{equipart}
\end{equation}
To reduce the mode temperature, we implement an electronic feedback cooling (Fig.~\ref{fig2}), or cold damping, technique \cite{squid2001}. The SQUID output voltage $V_o = A I_s$ is passed through a passive low-pass filter $D\left( \omega \right)$ with cut-off frequency at $200$ Hz, and the current $I_D=A D I_s$ is fed back to the signal circuit. The low-pass filter is used to phase-shift by $\pi$/2 the feedback current $I_D$, which is then proportional to the derivative of the oscillating current of the mode. As a consequence, the effect of the feedback current is equivalent to that of a viscous damping, similarly to the case of a mechanical oscillator subjected to a force proportional to its velocity. This additional damping can be represented by an equivalent resistor $R_D$ in series with $R_k$, which can be calculated from the model of Fig.~\ref{fig2}:
\begin{equation}
R_D = \Re \left( \frac{{iAD}}{{1 - AD}}\omega L_{{\rm in}} \right)
\end{equation}
In particular, in our experiment $R_D \cong |AD| \omega L_{\rm {in}}$, as $|AD|\ll 1$ and $AD$ is almost purely imaginary. Inductance and capacitance of the mode are not significantly modified, provided that the feedback current phase-shift is close to $\pi/2$. Therefore, we write the total impedance of the circuit under feedback cooling conditions, for $\omega \simeq \omega_{k}$, as $Z_k ^ \prime \left( \omega  \right) = R_k + R_D + i\omega L_k  + 1/\left( {i\omega C_k } \right)$. As customary, the relative strength of the feedback damping can be expressed by the ratio of the feedback to the intrinsic damping resistance (referred to the $k$-th mode) $g_k=R_D/R_k$. Then, the quality factor of the $k$-th mode under feedback cooling is reduced to $Q_k ^ \prime = Q_k/(1+g_k)$. 

According to the fluctuation-dissipation theorem, the thermal noise in the $k$-th normal mode is generated by a voltage noise source $V_{\rm{th}}$ with single-sided power spectral density $S_{V_{\rm{th}}}=4 k_B T_0 R_k$, where $T_0$ is the temperature of the thermal bath. This voltage noise is the effect of the interaction of the resonators with the microscopic degrees of freedom of the thermal bath, and therefore it is not affected by the feedback cooling. For $\omega \simeq  \omega_k$ the power spectrum of the current noise induced in the resonator is:
\begin{equation}
S_{I_{\rm{th}}}  = \frac{S_{V_{\rm{th}} }}{{\left| {Z_k ^ \prime} \right|^2 }} = \frac{{4k_B T_0 \omega _k }}{{Q_k L_k }}\frac{{\omega ^2 }}{{\left( {\omega ^2  - \omega _k ^2 } \right)^2  + \left( {{{\omega _k \omega } \mathord{\left/
 {\vphantom {{\omega _k \omega } {Q_k ^ \prime }}} \right.
 \kern-\nulldelimiterspace} {Q_k ^ \prime }}} \right)^2 }}.   \label{spettro}
\end{equation}
The Lorentzian shape of the spectrum is determined by the feedback-reduced quality factor $Q_k ^ \prime$, while the prefactor is determined by the intrinsic quality factor $Q_k$. Integration of Eq.~(\ref{spettro}) over frequency yields the total mean square current noise associated with mode $k$:
\begin{equation}
\left\langle {I_k ^2 } \right\rangle  = \frac{{k_B T_0}}{{L_k }}\frac{{Q_k ^\prime}}{{Q_k }} = \frac{{k_B T_0}}{{L_k }}\frac{1}{{1 + g_k }}, \label{iq}
\end{equation}
so that, similarly to the quality factor, the effective temperature of the mode under feedback cooling is reduced to:
\begin{equation}
T_k = \frac{T_0}{{1 + g_k }}.  \label{simple}
\end{equation}As the temperature reduction is determined by the $Q$-reduction, the larger is the initial quality factor, the larger is the achievable feedback cooling.

We can also rewrite Eq.~(\ref{spettro}) in terms of the mode temperature instead of the bath temperature:
\begin{equation}
S_{I_{\rm{th}}}  = \frac{{4k_B T_k \omega _k }}{{Q_k ^ \prime L_k }}\frac{{\omega ^2 }}{{\left( {\omega ^2  - \omega _k ^2 } \right)^2  + \left( {{{\omega _k \omega } \mathord{\left/
 {\vphantom {{\omega _k \omega } {Q_k ^\prime}}} \right.
 \kern-\nulldelimiterspace} {Q_k ^\prime}}} \right)^2 }}. \label{spettrobis}
\end{equation}
This is precisely the expected power spectrum of a passive resonator with quality factor $Q_k ^ \prime$ at thermal equilibrium at temperature $T_k$. Actually, the resonator is not at thermal equilibrium at $T_k$, but is rather in a dynamical equilibrium between the thermal bath at $T_0$, and the measuring-feedback system, which acts as a very low temperature bath. 

In the above analysis we have neglected the back-action and additive noise of the SQUID amplifier, which are expected to set a lower limit to the cooling efficiency. If the SQUID noise is taken into account, with the further simplified assumption of uncorrelated noise sources, the following refined expression of the mode temperature can be derived from the model in Fig.~\ref{fig2}:
\begin{equation}
T_k  = \frac{1}{1 + g_k }\left( T_0 +\frac{Q_k S_{V_n}}{4 k_B \omega_k L_k} \right)   + \frac{1}{{4k_B }}\frac{{g_k ^2 }}{{1 + g_k }}\frac{{\omega _k L_k }}{{Q_k }}S_{I_n } ,  \label{refined}
\end{equation}
where $S_{I_n}$ is the spectral density of the SQUID measurement noise $I_n$, fed back by the cooling loop, and $S_{V_n}$ is the spectral density of the SQUID back-action noise $V_n$. For our present setup, the back-action contribution is almost negligible with respect to thermal noise. According to Eq.~(\ref{refined}), there is an optimum value of $g_k$, for which the temperature achieves a minimum. This minimum achievable temperature is slightly dependent on the considered mode, and is of order 40 $\mu$K for our present setup. 

We point out that feedback cooling of the modes does not improve the sensitivity of the system as GW detector. In fact, the cooling is due to a modification of the effective response of the system to any kind of excitation. Therefore, it suppresses in the same way both the thermal noise and the external signal originated by an impinging GW.
\begin{figure}[!h]
\includegraphics{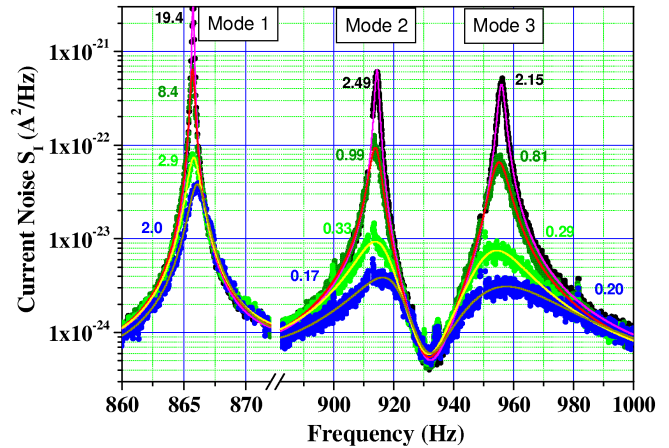}
\caption{(Color online) Power spectrum of the current noise in the normal modes, as measured by the SQUID amplifier. The noise spectra are related to four different feedback settings. Each noise spectrum is well-fitted by a proper combination of three Lorentzian curves. Each Lorentzian peak is labelled by the corresponding effective temperature, measured in mK.} \label{fig3}
\end{figure}

\begin{figure}[!h]
\includegraphics{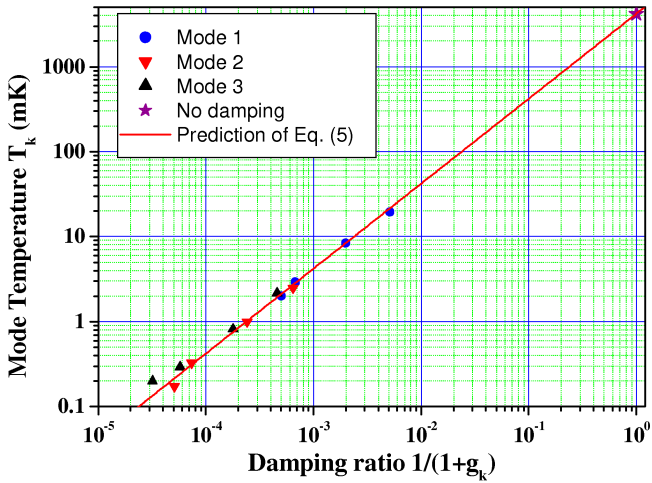}
\caption{(Color online) Effective temperature of the modes as function of the damping ratio. The straight line is the mode temperature predicted by Eq.~(\ref{simple}), with $T_0$ fixed to the value of the bath temperature $T_0=4.2$ K, and no free parameters. The theoretical limit at no feedback damping is also shown, to provide a graphical visualization of the achieved temperature reduction.} \label{fig4}
\end{figure}

To test the feedback cooling technique, we modified the standard operating conditions of the AURIGA detector, and set four different values of the feedback cooling gain $D$, corresponding to four values of the relative feedback damping $g_k$. For each setting, we measured the power spectral density of the current detected by the SQUID sensor, and averaged for a time of roughly one hour. The four spectra are shown in Fig.~\ref{fig3}. The overall power spectral density for a given setting can be accurately fitted by a proper combination of three Lorentzian curves, one for each mode. In fact, differently from previous experiments focused on a single mode, we simultaneously cool all three modes. The three-mode fitting curve is also shown in Fig.~\ref{fig3}, superimposed on the corresponding noise spectrum. For $\omega \simeq \omega_k$, the fitting function can be approximated by the single-mode expression Eq.~(\ref{spettrobis}), from which we can infer the effective temperature of each mode $T_k$. The experimental values of $T_k$ as function of the damping ratio $\left(1+g_k \right)^{-1}$ are shown in Fig.~\ref{fig4}. For a given feedback setting, different values of $g_k$ are associated to the three modes, because the intrinsic resistances $R_k$ are in general different. For instance, in our measurements $g_1$ ranges from $190$ to $2000$ and $g_3$ ranges from $2200$ to $30000$. 
The values of $g_k$ are still small enough to make almost negligible the effect of the SQUID noise. Therefore, we expect the effective temperatures to follow the simple behaviour described by Eq.~(\ref{simple}). The straight line in Fig.~\ref{fig4} represents $T_k$ for all $3$ modes, calculated using Eq.~(\ref{simple}) without free parameters, with $T_0$ fixed to the thermodynamic temperature of the bath, $T_0$=$4.2$ K. The data are in good agreement with the predictions of the model, even well below $1$ mK. The lowest achieved temperature are $T_1=2.0$ mK, $T_2=0.17$ mK and $T_3=0.20$ mK. The lowest temperatures for modes 2 and 3 correspond to an average occupation number $\left\langle {N_k} \right\rangle =k_B T_k / \hbar \omega_k \approx 4000$. 

These results represent an improvement by more than one order of magnitude with respect to the lowest temperature reported in literature for actively cooled macroscopic mechanical resonators \cite{poggio}, and only experiments performed at much higher frequency \cite{naik} have achieved lower occupation number. Although in our experiment we actively cool electromechanical normal modes rather than purely mechanical modes, we believe that the comparison is meaningful, as all three normal modes collect a significant fraction of the energy of the two mechanical resonators. For instance, from the detector calibration we estimate that the mode $2$, the coolest one, collects about 36\% of the energy deposited by an impulsive excitation in the bar resonator. Our cooling results are particularly relevant because of the enormous size of the elements of our system with respect to previously analysed micromechanical systems. This overturns the common belief that the cooling should be easier for lighter resonators. To explain this apparent paradox, we observe that the the relevant parameters in determining the cooling capability are the displacement sensitivity and the intrinsic quality factor of the resonator, which sets the potential $Q$-reduction ratio. In our case, the latter is one order of magnitude higher than that usually achieved by micromechanical resonators. In fact, according to a well-established empirical rule, the quality factor of a mechanical resonator scales roughly with the volume to surface ratio, suggesting a limiting factor in the surface dissipation mechanisms \cite{ekinci}. As a consequence, large resonators made of high $Q$ material, with linear dimensions in the $1$ cm –- $1$ m range, can easily reach quality factor as large as $10^6$–-$10^7$, whereas achieving $Q>10^5$ in micron-sized resonators has as yet proven difficult \cite{liu}. We also notice that a very large quality factor is a necessary condition for the observation of any macroscopical quantum behaviour. In fact, the timescale $\tau$ for decoherence in a resonator with quality factor $Q$ and temperature $T$ is proportional to to the ratio $Q/T$ \cite{corbitt,zurek}. 

Significant improvements with respect to the results presented here are expected by performing a dedicated experiment on a mid-scale ($0.01$--$1$ kg) resonator at subkelvin bath temperature. Conventional dilution refrigerators can be used to cool kg-scale masses down to $10$ mK \cite{CUORE}. At the same time, resonators made of very low dissipation material, like silicon or sapphire, can reach quality factors as large as $10^9$ \cite{mcguigan}. Moreover, our capacitive-SQUID measurement system, similarly to SET-based ones \cite{naik}, is naturally compatible with ultralow temperatures, due to the very low power dissipation, of order $10^{-10}$ W for typical dc SQUIDs. Eventually, the maximum cooling ratio will be limited by the SQUID sensitivity, according to Eq.~(\ref{refined}). As state-of-art devices can approach the quantum limit \cite{carelli,muck,falferi}, resonator temperatures lower than $1$ $\mu$K are achievable at kHz frequency, corresponding to a single-digit occupation number of the quantum oscillator.

\begin{acknowledgments}
We thank W. J. Weber for discussions and reading of the manuscript
\end{acknowledgments}

\end{document}